\documentclass[
	article,
	12pt,				
	oneside,
	a4paper,			
	english,			
	french,				
	spanish,			
	brazil,				
	]{abntex2}

\usepackage[T1]{fontenc}		
\usepackage[utf8]{inputenc}		
\usepackage{indentfirst}		
\usepackage{color}				
\usepackage{graphicx}			
\usepackage{microtype} 			

\usepackage{lipsum}				
\usepackage[brazilian,hyperpageref]{backref}	 
\usepackage{abntex2cite}	
\usepackage{cite}

\usepackage{amsmath}
\usepackage{array, amsfonts}
\usepackage{amsthm}
\usepackage{amssymb}
\usepackage{enumerate}

\usepackage{mathrsfs}
\usepackage{stackengine}
\usepackage{bookmark}
\usepackage{enumitem}

\usepackage{mathtools}
\usepackage{todonotes}
\usepackage{graphicx}
\usepackage{multicol}
\usepackage{caption}
\usepackage{subcaption}
\usepackage{pdfpages}
\usepackage[a4paper,left=3cm,right=3cm,top=2.5cm,bottom=2.5cm]{geometry}

\usepackage{times}

\usepackage{microtype}

\usepackage{lineno}

\titulo{Emergence of complex data \\ from simple local rules\\ 
	in a network game\thanks{
		Authors acknowledge the partial support from CNPq through their individual grants: F. S. Abrahão (313.043/2016-7), K. Wehmuth (312599/2016-1), and A. Ziviani (308.729/2015-3). Authors acknowledge the INCT in Data Science – INCT-CiD (CNPq 465.560/2014-8). Authors also acknowledge the partial support from FAPESP (2015/24493-1), and FAPERJ (E-26/203.046/2017). We also thank Hector Zenil, Mikhail Prokopenko, \'{I}tala M. Loffredo D'Ottaviano, Leonardo Lana de Carvalho, and Andréa Naccache for suggestions and directions on related topics investigated in this article.
		}}
\tituloestrangeiro{}

\autor{Felipe S. Abrah\~{a}o\thanksmark{2}   \\ {\small email: \textit{fsa@lncc.br}} 
	\and Klaus Wehmuth\thanksmark{2}  \\ {\small email: \textit{klaus@lncc.br}}
	\and Artur Ziviani\thanks{National Laboratory for Scientific Computing (LNCC) -- 25651-075 – Petropolis, RJ -- Brazil.} \\ {\small email: \textit{ziviani@lncc.br}}}

\local{Brasil}
\data{2020}


\SingleSpacing

\makeindex

\theoremstyle{definition}
\newtheorem{definition}{Defintion}[section]

\theoremstyle{remark}

\theoremstyle{remark}

\theoremstyle{definition}

\theoremstyle{remark}

\theoremstyle{remark}

\begin{document}

\selectlanguage{english}

\frenchspacing 




\maketitle



\begin{resumoumacoluna}[Abstract]
	As one of the main subjects of investigation in data science, network science has been demonstrated a wide range of applications to real-world networks analysis and modeling. For example, the pervasive presence of structural or topological characteristics, such as the small-world phenomenon, small-diameter, scale-free properties, or fat-tailed degree distribution were one of the underlying pillars fostering the study of complex networks. Relating these phenomena with other emergent properties in complex systems became a subject of central importance. By introducing new implications on the interface between data science and complex systems science with the purpose of tackling some of these issues, in this article we present a model for a network game played by complex networks in which nodes are computable systems. In particular, we present and discuss how some network topological properties and simple local communication rules are able to generate a phase transition with respect to the emergence of incompressible data.
	
\end{resumoumacoluna}



%

%


\textual



\section[Introduction]{Introduction}\label{sectionIntro}

Computation, information, and networks are three concepts that are of major importance in the contemporary world, where the social impacts of pervasive network dynamics in our digital life, big data, and the increasing power of data analysis bridge the gap between complex systems science and the everyday dynamics of our lives.
Long have been a common sense
that natural systems can be understood as being organized as networks of many interacting
units \cite{Lewis2009}.
For example, interacting molecules in living cells, nerve cells in the brain, computers in a telecommunication network, and socially interacting people. 
It is intuitive to break a system into two constitutive realms: that of individual 
components, e.g., the laws an atom is subjected to, how a real-world computer works, or how a human being thinks or behaves;
and that of the nature of the connections (or interactions), e.g., the internet communication protocols, the vehicles' characteristics in a transportation network, or the type of human friendships. 

However, there is a third realm, sometimes overlooked, that is also important in order to determine the system's functioning: the realm of patterns of connections \cite{Newman2010}.
In other words, beyond the mere fact that a system is composed of parts and that these parts are working in interaction, the patterns, structures or topological properties of a network may play a significant---if not dominant---role in the dynamics of the entire system.
Indeed,
recent advances in complex network theory indicate that this third notion may be more than just a representation scheme or metaphor \cite{Barabasi2016,Lewis2009}.  
Triggered by the availability of large amounts of data on large real-world networks, combined with fast computer power even on scientists’ desktops, the field is reaching consensual saturation point, being called by the umbrella term \emph{network science} \cite{Zarate2019nat,Barabasi2016}, and plays a central role in data science in general. 
Applications ranges from internet communication protocols, epidemics, prevention of computer viruses, fail-safe computer networks engineering, regulatory
circuits of the genome, and ecosystems \cite{Barabasi2016}.

Rooted in graph theory, e.g., from Euler's work to Erdős–Rényi (ER) random graphs, the investigation of complex networks highlights the pervasive presence of heterogeneous structural characteristics of real-world networks \cite{Barabasi2016}.
This is the case of the \emph{small-world effect} \cite{Lewis2009}, where the average shortest path distance, or mean geodesic distance, between any pair of vertices increases up to a logarithmic term of the network size.
While popularly known from the ``six degrees of separation'' phenomenon in social science, the \emph{small-world network} gained a more formal mathematical ground after the Watts–Strogatz model in which, in addition to the short mean geodesic distance, the generated networks have e.g. a high clustering coefficient (i.e., the tendency of vertex neighbors to be connected to other neighbors) \cite{Lewis2009}.
Regarding the heterogeneity of vertex degrees, another commonly found characteristic is a fat-tailed (or heavy-tailed) distribution, for example when the vertex degree distribution follows a power-law, as in the Barabási-Albert (BA) model (aka scale-free networks), and not a Poisson distribution like in the traditional ER model \cite{Barabasi2016}.

This way, in consonance with the pursuit of a theory for evolutionary, computational, dynamical, and informational aspects in complex systems \cite{Mitchell2009}, the study of general and unifying models for the emergence of complexity and network topological properties keeps attracting the interest of the researchers of network science, data science, and complex systems science 
\cite{Michail2018}.
In this direction, 
information-theoretic approaches have been demonstrating fundamental contributions with the purpose of defining, detecting, or modeling the presence of systemic properties, such as emergence, complexity, and self-organization, in systems with stochastic dynamics \cite{Prokopenko2009}.
Moreover, not only in computable systems, but also as refinements of more traditional statistical approaches, recent advances have been highlighting the algorithmic-informational perspective and showing new fundamental results on: open-endedness and evolutionary systems \cite{Chaitin2012,Hernandez-Orozco2018,Hernandez-Orozco2018a}; network complexity \cite{Zenil2018a,Abrahao2018dextendedarxiv2020}; machine learning and causality \cite{Zenil2019}; cybernetics and control theory \cite{Zenil2019d,Zenil2019c}; and emergence of complexity in networked systems \cite{Abrahao2017publishednat,Abrahao2018publishednat}.
Following this latter approach, we present in this article an investigation of network topological conditions that trigger a phase transition in which algorithmic networks eventually begin to produce an unlimited amount of average emergent algorithmic complexity as the population size grows toward infinity.
These topological conditions can be any property that reflects a strong diffusion power through the network, such as the small-diameter phenomenon \cite{Abrahao2017publishednat} or a classical case of scale-free network \cite{Abrahao2018publishednat}.
Within the context of networked computable systems, we demonstrate the existence of emergence that is proved to be irreducible to its individual parts, universal, and independent of any arbitrarily fixed observer.

\section{A model for networked computable systems}\label{sectionModel}

In this section, we present the general mathematical model for the study of networked machines, which can share information with each other across their respective network while performing their computations. 
The model is defined in a general sense in order to allow future variations, to add specificities and to extend the model presented, while still being able to formally grasp a mathematical analysis of systemic features like the emergence of information and complexity along with its related phenomena and, this way, proving theorems.
It was introduced in \cite{Abrahao2016bnat,Abrahao2017publishednat} and we have studied other variations of this first model with a static scale-free network topology in \cite{Abrahao2018publishednat} and with a modified communication protocol to synergistically solve mathematical problems in \cite{Abrahao2019nat}.
In the present article we will focus on the model in \cite{Abrahao2017publishednat}.

The main idea behind the general model is that a population of formal theoretical machines can use communication channels over the graph's edges. 
Thus, the graph topology causes this population to be networked. 
Once the elements of the population start to exchange information, it forms an overarching model for a system composed of interacting subsystems.
Following this general approach, one can understand these mathematical models as a merger of algorithmic (and statistical) information theory and complex networks, while theoretically combining fundamental notions from distributed computing, multiagent systems, adaptive complex systems, game theory, and evolutionary biology.
We refer to such models as \emph{algorithmic networks} \cite{Abrahao2016bnat,Abrahao2017publishednat}. 
So, algorithmic networks are networks of algorithms in the precise sense where the nodes of the network are computable systems.
For the present purposes, one may consider each node as a program of a universal Turing machine, which justifies calling either the nodes or the elements of the population of an algorithmic network as \emph{nodes/programs}.

Aiming at a wider range of different network configurations, we ground our formalism on multiaspects graphs (MAG) as presented in \cite{Wehmuth2016b}. 
In this way, one can mathematically represent extra aspects or dimensions that could appear in complex networks. It has been shown that the MAG abstraction enables one to formally represent and computationally analyze networks with additional representational structures.
For having additional dimensions in which the nodes belong (or are ascribed to), e.g., time instants or layers, such networks are called \emph{multidimensional networks} (or high-order networks): for example,
dynamic  (i.e., time-varying) networks \cite{Costa2015a,Abrahao2018dextendedarxiv2020}, multilayer networks \cite{Kivela2014}, and dynamic multilayer networks \cite{Wehmuth2018}.  
Moreover, MAG abstraction facilitates network analysis by showing that their aspects can be isomorphically mapped into a classical graphs \cite{Wehmuth2016b}.

In a broad sense, one can think of an algorithmic network as a theoretical multidimensional network-distributed computing model in which each node (or vertex) computes using the shared information through the network.
The computation of each node may be seen in a combined point of view or taken as individuals. Respectively, nodes/programs may be computing using network's shared information to solve a common purpose \cite{Abrahao2019nat}---as the classical approach in distributed computing---or, for example, nodes may be ``competing'' with each other---as in a game-theoretical perspective, which we employ in this article (see Section~\ref{sectionBBIG}).
For the present purposes, we are interested in the average fitness (or payoff), and its related emergent complexity that may arise from a process that increases the average fitness. 

\begin{definition}\label{BdefAN}
	We define an \emph{algorithmic network} $ \mathfrak{N} = (\mathscr{G}, \mathfrak{P}, b)$ upon a population of theoretical machines $\mathfrak{P}$, a multiaspect graph $\mathscr{G}=(\mathscr{A},\mathscr{E})$ and a function $b$ that causes aspects of $\mathscr{G}$ to be mapped into properties of $\mathfrak{ P }$, so that a vertex in $\mathrm{V}(\mathscr{G}) $ corresponds one-to-one to a theoretical machine in $\mathfrak{ P }$ and the communication channels through which nodes can send or receive information from its neighbors are defined precisely by (composite) edges in
	$ \mathscr{G} $.  		
\end{definition}

The MAG $\mathscr{G}$, as previously defined in \cite{Wehmuth2016b}, is directly analogous to a graph, but replacing each vertex by a $n$-tuple, which is called the composite vertex.
Note that a graph is a particular case of a MAG that has only one aspect (i.e., only one node dimension).
A \emph{population} $\mathfrak{P}$ is a sequence (or multiset) with elements taken from $L$ in which repetitions are allowed, where $L$ is the language on which the chosen theoretical machine are running.
A \emph{communication channel} between a pair of elements from $\mathfrak{P}$ is defined in $\mathscr{E}$ by a composite edge (whether directed or not) linking this pair of nodes/programs.
A directed composite edge (or arrow) determines which node/program sends an output to 
another node/program, which in turn takes this information as input. An undirected composite edge 
(or line) may be interpreted as two opposing arrows. 
We say an element $ o_i \in \mathfrak{P} $ is \emph{networked} \textit{iff} there is $ 
\mathfrak{N} $ such that $o_i$ is running as a node of $ \mathfrak{N} 
 $, where $ \mathscr{E} $ is non-empty. 
That is, there must be at least one composite edge connecting two elements of the algorithmic network.
We say $o_i$ is \emph{isolated} otherwise.
We say that an input $ w \in L $ is a \emph{network input} \textit{iff} it is the only external source of information every node/program receives and it is given to every node/program before the algorithmic network begins any computation. 
A \emph{node cycle} in a population $ \mathfrak{ P } $ is defined as a node/program returning an output, which, in the particular studied model \cite{Abrahao2017publishednat} described in Section~\ref{sectionBBIG}, is equivalent to a node completing a halting computation.
If this node cycle is \emph{not} the last node cycle, then its respective output is 
called a \emph{partial output}, and this partial output is shared (or not, which depends on 
whether the population is networked or isolated) with the node's neighbors, accordingly 
to a specific information-sharing protocol (if any).
On the other hand, if the node cycle is the last one, then its output is called a \emph{final output}.


Our formalism enables one to represent a wide range of variations of algorithmic networks with the purpose of modeling a particular problem that may arise from a networked complex system. 
For example, the networked population may be synchronous or asynchronous, have a set of information-sharing strategy or none, a randomly generated population or a fixed one, with communication costs or without them, etc. 
In addition, the network topology that determines the communication channels may be dynamical, with weighted edges, multilayer etc. 
In particular, all models considered hereafter, as described in Section~\ref{sectionBBIG}, are synchronous (i.e., there are communication rounds that every node must respect at the same time), have a fixed information-sharing strategy (i.e., a communication protocol), have a randomly generated population of programs, and no communication cost is considered.

\section{Local fitness optimization in the Busy Beaver imitation game}\label{sectionBBIG}

 
Now, we explain a particular case of algorithmic network defined by a very simple local rule (i.e., a rule each node follows with respect to its immediate neighbors) that optimizes the fitness value of each node individually.
Then, later on in this article, we will discuss the impacts on the global behavior of the algorithmic network that this simple rule of communication produces.

The main idea of the model in \cite{Abrahao2017publishednat} is as follows: 
take a randomly generated set of programs;
they are linked, constituting a dynamic network which is represented by a time-varying graph (or a multiaspect graph with two aspects);
each node/program is trying to return the ``best solution'' it can; 
and eventually one of these nodes/programs end up being generated carrying beforehand a ``best solution'' for the problem in question; 
this ``best solution'' is spread through the network by a diffusion process in which each node is limited to only imitate the fittest neighbor if, and only if, its shared information is ``better'' than what the very node can produce (see the imitation-of-the-fittest protocol below). 

Indeed, a possible interpretation of the diffusion described to the above is \emph{average optimization through diffusion} in a random sampling. 
Whereas optimization through selection in a random sampling may refer e.g. to evolutionary computation or genetic algorithms, optimization here is obtained in our model in a manner that a best solution also eventually appears, but is diffused over time in order to make every individual as averagely closer to the best solution as they can. 
Therefore, the underlying goal of this process would be to optimize the average fitness of the population by expending the least amount of diffusion time (or communication rounds).

As in \cite{Chaitin2012,Abrahao2015}, we use the \emph{Busy Beaver function} $ BB(N) $ as our complexity measure of \emph{fitness}. 
A function $ BB(N) $, where $ BB : \, \mathbb{N} \to \, \mathbb{N} $, returns the largest integer that a program $ p \in \mathbf{L_U}  $ with length $ \leq N $ can output.
Naming larger integers relates directly to increasing algorithmic complexity \cite{Chaitin2012}. 
Thus, the ``best solution'' assumes a formal interpretation of fittest final output (or payoff). 
The choice of the word ``solution'' for naming larger integers now strictly means a solution for the Busy Beaver problem. 
Also note that several uncomputable problems are equivalently reduced to the Busy Beaver one, including the halting problem.
In addition, the Busy Beaver function offers other immediate advantages in measuring the complexity of the fitness value.
For example, it grows faster than any computable function, while being scalable (i.e., every fitness value alone can be eventually reached by some individual computable system);
integers being fitness values is universal with respect to Turing machines, while the values themselves are totally dependent on the nodes' initial conditions or context;
the value of $ BB(N) $ is incompressible, i.e., an arbitrary universal Turing machine needs at least, except for a constant, $N$ bits of information to calculate the value of $ BB(N) $.
This way, with a fixed fitness function that works as a universal parameter for every node/program's final (and partial) output, it makes sense to have an interpretation of these running algorithmic networks in \cite{Abrahao2017publishednat} as playing a \emph{networked Busy Beaver game}: 
during the node cycles, each node is trying to use the information shared by its neighbors to return the largest integer it can. 
The larger the final output integer, the better the payoff (or fitness).

We employ the term \emph{protocol} as an abstraction of its usage in distributed computing and telecommunications. 
A protocol is understood as a set of rules or algorithmic procedures that nodes must follow at the end of each node cycle when communicating. 
It can be seen as the strategy or ``rules'' for communications under a game-theoretical perspective, and 
within this context an algorithmic network can be interpreted as playing a game in which each 
node is trying to return the ``best solution'', or the best fitness value, it can.
In our studied model, we want to investigate one of the 
simplest, computationally cheapest, or ``worst'' ways that networked nodes can take advantage of its neighbors' information sharing and compare with the best that isolated nodes can do alone.
Hence, we oblige the networked nodes to follow the \emph{imitation-of-the-fittest protocol} (IFP), which is a decidable procedure in which a networked node compares its neighbors' partial outputs and propagates the program of the neighbor that have output the largest integer.
But it only does so if, and only if, this integer is larger than the one that the very node has output in first place. 
This way, the networked population is in fact limited to simple imitation:
it is a game with a single strategy, i.e., the IFP, if the node is networked; 
and a single strategy, i.e., doing the best (without any specified protocol or strategy) the node can alone, if the population is not networked.
Therefore, we say such an algorithmic network is playing a \emph{Busy Beaver imitation game} (BBIG) \cite{Abrahao2017publishednat}.

\section{Expected emergent open-endedness from universal complexity measures}\label{sectionEOE}

Now, the question is: 
how much more algorithmic complexity can this diffusion process generate on the average compared with the best nodes/programs could do if isolated?
Toward an answer to this question, a comparison between the algorithmic complexity of what a node/program can do when networked and the algorithmic complexity of the best a node/program can do when isolated gives the \emph{emergent algorithmic complexity} of the algorithmic network. 
Instead of asking about how much complexity is gained by systems over time, as in evolutionary biology and artificial life, we are focusing on another akin question: 
how much complexity is gained by systems when the number of parts increases? 
Or, more specifically in our case, how much more emergent algorithmic complexity arises on the average when the number of nodes increases?

Once we are restricted to only dealing with networked computable systems, the functioning of these systems occurs in a totally deterministic way.
And more than that, they are computable, i.e., for any one of them there is a Turing machine that, given the environmental conditions or context as input, can always completely determine their next behavior or state from a previous behavior or state.
In this way, algorithmic information theory (AIT) sets foundational results from which one directly obtains an irreducible information content measure \cite{Chaitin2004} of a mathematical object being generated by a computable process; 
this object may be e.g. the output of a machine or the future state of a computable system.
More precisely, the quantification of irreducible information content can be stated in bits and is given by the (unconditional) \emph{algorithmic complexity} of an object $x$, i.e., the length of the shortest program that outputs  $x$ when this program is running on an arbitrarily chosen universal Turing machine.

In addition, algorithmic complexity is a quantity that is invariant---therefore, irreducible or incompressible---for any other computable process that can generate $x$, except for an additive constant: that is, the two quantities of complexity can only differ by an additive constant for any $x$ and this constant only depends on the choice of the machine and the computable process, so that the constant does not depend on $x$.
On the other hand, algorithmic complexity is an optimal information content measure.
In other words, one can also show that there is a universally maximal recursively enumerable probability semimeasure $\mu$ for the space of all encoded objects such that the time-asymptotic approximation to the probability $\mu(x)$ of occurrence of $x$ is always larger than (except for a multiplicative constant) any other time-asymptotic approximation to the probability $\mu'(x)$ of occurrence of $x$.
More formally, for any recursively enumerable probability semimeasure $\mu'$ for the space of all encoded objects, there is a multiplicative constant $c$, which does not depend on the object $x$, such that, for every $x$, one has that
$ c \, \mu(x) \geq \mu'(x) $ holds. 
And this result holds even if one has zero knowledge about the actual probability of occurrence of $x$.
Indeed, one can already note that such zero-knowledge characteristic differs from traditional statistical inference methods, where it is in general assumed that the stochastic random source is, at least, stationary and ergodic.

As one of the main and most profound results in AIT, the \emph{algorithmic coding theorem}, one can show that the probability of $x$ be generated by any possible randomly generated (prefix) Turing machine, the above universally maximal probability semimeasure $\mu(x)$, and the probability of occurrence of the shortest program that generates $x$ are in fact three equivalent values, except for a multiplicative constant that does not depend on $x$.
Thus, at least for the realm of deterministic computable processes, algorithmic complexity is a measure of information content that is irreducible/incompressible and universal, in the sense that it is invariant on the choice of the object at stake and any computable process of measuring the irreducible information content of $x$ equivalently agrees (up to object-independent constant) about the value.
It is a mathematically proven ``bias toward simplicity'' for the space of all generative computable processes.
Not only for the unconditional form of algorithmic complexity, the same phenomenon also holds for the conditional algorithmic complexity, i.e., the length of the shortest program that generates $y$ given the input $x$.
This way, algorithmic complexity appears as an auspicious mathematical form of information content measure, specially for those computable systems whose behavior is dependent on the information received from the environment:
the algorithmic complexity of $y$ given $x$ is a value that is, at the same time, totally dependent on the input (i.e., the initial conditions or previous context), irreducible, and universal.
Therefore, as desirable, quantifying an emergence of complexity in computable systems from a direct comparison between the algorithmic complexity of the networked/interacting case (i.e., $y$) and the isolated case (i.e., $x$) gives a value that is irreducible and universal, although might vary only if the system's environment in which this comparison took place changes.

We follow a consensual abstract notion of \emph{emergence}
\cite{DOttaviano2004,Prokopenko2009}
as a systemic feature or property that appears only if the system is analyzed (theoretically or empirically) as a ``whole''. 
Thus, the algorithmic complexity (i.e., an irreducible number of bits of information) of a node/program's final output when networked\footnote{ That is, interacting with other parts of the system.} minus the algorithmic complexity of a node/program's final output when isolated formally defines an irreducible quantity of information that \emph{emerges} with respect to a node/program that belongs to an algorithmic network. 
We call it as \emph{emergent algorithmic complexity} (EAC) of a node/program \cite{Abrahao2017publishednat}. 
Consequentially, note that if a system is analyzed as a separated\footnote{ The subparts do not need to be necessarily apart from each other, but each part in this case would be taken as an object of investigation where no information enters or exits anyway.} collection of ``subparts'', the EAC of a node/program will be always $0$.
Note that this quantity of bits may be $0$ or negative in some cases.
Therefore, this measure of emergent algorithmic complexity may also be suitable for measuring the cases where algorithmic complexity was ``lost'' when the system is networked. 
We leave the study of such degenerate cases as an important future research and, for the present purposes, we are only interested in the situations in which EAC is positive.

A distinction is crucial: the EAC of a node/program must be not confused with the EAC of the \emph{entire} algorithmic network. Measuring the emergent algorithmic complexity of the algorithmic network taking into account every node/program ``at the same time'' is---as our intuition demands to be---mathematically different from looking at each individual final output's algorithmic complexity. For example, one may consider the algorithmic information of each node/program combined (in a non-trivial way) with the algorithmic information of the network's topology.
This relies upon the same distinction between the joint algorithmic complexity of $x$ and $y$ and the algorithmic complexity of each one taken separately.
The sum may not always match the joint case \cite{Chaitin2004}.  
Within the framework of algorithmic networks, this ``whole'' emergent algorithmic complexity compared with each individual node can be formally captured by the \emph{joint} algorithmic complexity of each node/program's final output when networked minus the \emph{joint} algorithmic complexity of each node/program's final output when isolated. 
That is, the algorithmic complexity of the networked population's output as a whole minus the algorithmic complexity of the isolated population's output as a whole. 
An initial step in the direction of tackling this problem is already mentioned in \cite{Abrahao2019nat}. 
Analyzing this systemic property is not part of the scope of the present article and it will be a necessary future research, not only in the context of networked computable systems, but it is also an open problem for multivariate stochastic processes \cite{Lizier2018}.

Thus, instead of investigating the \emph{joint} or \emph{global} EAC of an algorihtmic network, one may look for a mean value of EAC for all nodes/programs. 
That is, we are focusing the \emph{local} EAC. 
The \emph{average} (local) \emph{emergent algorithmic complexity} of a node/program (AEAC) is defined by the mean on all nodes/programs' (and possible network's topologies) EAC. 
It gives the average emergent complexity of the nodes/programs' respective fitnesses (or, in a game-theoretical interpretation, payoffs) in a networked population, once there is a fitness function that evaluates final outputs.
Larger positive values of AEAC mean that a node/program needs more irreducible information on the average than it already contains, should it try to compute isolated what it does networked. 
A system with a larger AEAC ``informs'' or ``adds'' more information to its parts on the average.

As the model described in Section~\ref{sectionBBIG} is an algorithmic network in which the population of machines is randomly generated from a stochastic process of independent and identically distributed (i.i.d) random variables under a self-delimiting program-size probability distribution, we can refer to the average EAC as \emph{expected} emergent algorithmic complexity (EEAC).
Therefore, both terms, \emph{average or expected}, can be used interchangeably hereafter.
Note here that, whereas the initial network input is completely arbitrary and the algorithmic network itself in the model described in Section~\ref{sectionBBIG} is a deterministic and computable distributed system (once the population of nodes/program is given), the initial generation of the nodes/programs of each algorithmic network is given by a stochastic i.i.d. process.
Thus, each of these algorithmic networks are deterministic (computable) processes, while the infinite process that results from increasing the size of the algorithmic networks and running them is a mixed process (i.e., partially deterministic and partially stochastic).   

Another important concept that came from complex systems science, specially from artificial life and evolutionary computation, is \emph{open-endedness}. 
It is commonly defined in evolutionary computation and evolutionary biology as the inherent potential of a evolutionary process to trigger an endless increase of distinct systemic behavior capabilities
\cite{Adams2017,Hernandez-Orozco2018}.
Thus, if an infinite space of distinct computable capabilities is eventually covered, this will necessarily lead to an unbounded increase of algorithmic complexity \cite{Hernandez-Orozco2018}.
This means that, in the long run, it will eventually appear an organism that is as complex as one may want.
Given a certain complexity value as target, one would just need to wait a while in order to appear an organism with a larger complexity than (or equal to) the target value---no matter how large this value is.
In turn, this implies that an infinite number of different organisms tends to appear in the evolutionary path after an infinite amount of successive mutations, bringing us equivalently back to the initial definition of open-endedness.

In fact, within the framework of \emph{metabiology}, as shown in \cite{Chaitin2012,Abrahao2015,Chaitin2018}, there is a cumulative\footnote{ Which allows organisms to recall its predecessors. } evolution model that reaches $N$ bits of algorithmic complexity after---realistic fast---$ \mathbf{ O }( N^2 ( \log(N) )^2 ) $ successive algorithmic mutations on one organism at the time---whether your organisms are computable \cite{Chaitin2012}, sub-computable \cite{Abrahao2015,Abrahao2016} or hyper-computable \cite{Abrahao2015}.
Metabiology is a transdisciplinary field based on evolutionary biology and algorithmic information theory that proposes a metatheoretical approach to the open-ended evolution of computable systems \cite{Chaitin2018,Chaitin2014nat}.  
Moreover, it is shown in \cite{Hernandez-Orozco2018} and experimentally supported in \cite{Hernandez-Orozco2018a} that the model introduced in \cite{Chaitin2012} satisfies the requirements for \emph{strong} open-ended evolution.
Thus, if one is restricted to the case of evolutionary computation in general\footnote{ That is, taking into account not only those with bounded computational resources, but also those with unbounded computational resources (like Turing machines).} computable systems, open-endedness is then stricly related to an endless increase of algorithmic complexity or irreducible information. 
And, since we are studying networked computable systems, we follow this algorithmic and universal approach to open-endedness in which undecidability and irreducibility plays a central role \cite{Hernandez-Orozco2018}.

What we have found is that, within the theory of algorithmic networks, open-endedness also appears in a similar fashion.
However, it emerges as an akin, but formally distinct, phenomenon to open-ended evolution (OEE): instead of achieving an unbounded quantity of algorithmic complexity over time (or successive mutations), an unbounded quantity of emergent algorithmic complexity is achieved as the population size increases indefinitely. 
Since it is a property that emerges depending on the amount of parts of a system only when these parts are interacting somehow (e.g., exchanging information) and this new quantity of algorithmic complexity/information is irreducible/incompressible with respect to the programs that governs the functioning of the respective isolated parts, this unbounded increase of EAC arises, by definition, as an emergent property. 
So, we refer to it as \emph{emergent open-endedness} (EOE) \cite{Abrahao2017publishednat}.
As discussed before, since we are dealing only with the local EAC and not with the global (or joint) EAC, then a more accurate term would be \emph{local emergent open-endedness}. 
For the sake of simplifying our nomenclature, we choose to omit the term ``local'' in this article.
Furthermore, in the case of an increase in the average EAC for every node/program, we refer to it as \emph{average (local) emergent open-endedness} (AEOE). 
And, since the population is randomly generated, we refer to AEOE as \emph{expected (local) emergent open-endedness} (EEOE).


We showed in \cite{Abrahao2017publishednat}
that there are network topological conditions and simple communication protocols that trigger EEOE as the randomly generated populations grows toward infinity.
In particular, a model of algorithmic networks for which we proved that it occurs is the one described in Section~\ref{sectionBBIG}; and the network topological conditions can be a strong diffusion power, so that larger fractions of the network are quickly covered by any signal spread by any node, or the presence of the small-diameter phenomenon, which guarantees that the entire network is covered under a small amount of hops, steps, or (in the case of synchronous algorithmic networks) communication rounds.
As shown in \cite{Abrahao2017publishednat},
these conditions caused the EEAC to increase as one may want, should the population size increases sufficiently.
And this occurs even if, for an arbitrarily large (but finite) population size, the EEAC is $0$ or negative.
The networked ``side of the equation'' of the EAC relies only on the simple imitation of the fittest neighbor, while the ``isolated side'' is free of any strategies or protocol so that each node can perform/compute without any restriction.
Thus, we are estimating the emergent algorithmic complexity that arises from a ``worst'' networked case compared with the ``best'' isolated nodes can do alone.
So, if in this worst-case scenario the EAC has increasingly positive integer values, then the EEAC (which is an average-case scenario lower bounded by the worst case) will behave the same way.
More precisely, the expected emergent open-endedness phenomenon tells us that, for large enough population sizes, the probability that these algorithmic networks have a larger AEAC tends to $1$.
The main idea behind the proof is that, given that such conditions are satisfied, there will be a trade-off between the number of communication rounds and the average density of networked nodes with the maximum fitness, so that there is an optimum balance between these two quantities in which, if a large enough average density of these nodes is achieved in a sufficiently small number of communication rounds, then EEOE is triggered.

\section{Emergence of unpredictable and irreducible data: \\ discussion, open problems, and future work}

EEOE is in fact a phenomenon that reflects a phase transition of complexity, in particular, an emergence of algorithmic complexity, with deep implications to the investigation of networked complex systems or any distributed processing of data:
for example, either for designing or engineering artificial computer networks; or analyzing real-world networks of complex system in which each node represents a system that is capable (allegedly) of performing some kind of computation, e.g., biological organisms or humans.

Note that our results show the existence of a phase transition in which, for a critical stage (in the case, a large enough population), the network will change its networked behavior (in comparison to the isolated one) so drastically that it will be impossible for any of the nodes/programs to compute (or computably predict) its own networked behavior.
This is the reason we call this transition as an \emph{expected emergent complexity phase transition}: an algorithmic complexity phase transition that is guaranteed to occur in the asymptotic limit, giving rise to the emergence of irreducible information solely by the fact the population of nodes/programs is networked.
In the case fitness (or payoff) is somehow connected to the complexity of the player's strategy, algorithmic networks theory is a theoretical model for future investigation of game-theoretical consequences of randomly generated arbitrary computable strategies for players without interaction in comparison to networked players' strategies.

Now, take for example real-world networks, such as ecosystems or human societies, where each element is an information processing system \cite{Mitchell2009,Prokopenko2009,Michail2018} that can send and receive information from each other.
Remember that the studied communication protocol in Section~\ref{sectionBBIG} is in fact one of the ``worst'' local rules of individual behavior that is capable of increasing the fitness with respect to its neighbors.
Then, assume for a moment that those real-world networks are composed of nodes/systems with a high enough computational power---indeed, a plausible suposition at least for nodes representing human beings---, so that they eventually begin to perform better than their neighbors in terms of an arbitrarily chosen fitness measure (which may assume unbounded, but reachable, values).
In addition, also assume the entire network is embedded into an ``environment'' that is capable of always ascribing fitness values to nodes.
Thus, now we know there are some network topological conditions, e.g., a strong diffusion power or the small diameter, that eventually enable some algorithmic networks to reach a phase transition point in which EEOE is triggered.

From a computational analysis perspective, EEOE immediately implies that, although graph-topological, structural, or connection-pattern modeling (or predictions) could be made by computational methods for network analysis and machine learning, modeling or predictions by artificial intelligence would be eventually unattainable or intractable with respect to the information content processed by the nodes.
This may be a desirable property for computer networks design, if one is aiming at the networked information processing being relatively uncomputable, or encrypted, to isolated nodes.
Moreover, if one is trying to take advantage of the network-distributed computation with the purpose of computing problems at a higher computational class, the EEOE phenomenon could be harvested from synergistic variations of the communication protocols.
This mathematical phenomenon may be also fruitful for explaining
synergistic behavior found in Nature and societies and
why some network topological properties seem to be favored in biological networks.
Indeed, algorithmic synergy was already shown to exist in networked resource-unbounded computable systems with a slight modification into the IFP \cite{Abrahao2019nat}.
Future research in this direction will be interesting for developing resource-bounded versions and, therefore, more realistic network-distributed computing models and architectures.

On the other hand, EEOE may be a property one is avoiding in order to keep the computer network processing power under control or below a certain degree of complexity.
Such an emergent phenomenon would impose a necessary limit for data analysis in those networks displaying EEOE, if the computational power of the observer is at the same level of the nodes---therefore, also including the case where the observer is one of the nodes.
For any arbitrarily chosen formal theory, or computer program, that an external observer chooses as framework, there will be a critical stage in which the network displays EEOE and any attempt to predict the networked behavior of the nodes will start to be relatively uncomputable (i.e., belonging to a higher level at a computational hierarchy).
In particular, as one can directly obtain from algorithmic information theory (AIT) \cite{Chaitin2004}, the networked behavior will be unpredictable in precise terms of an increasing quantity of bits that are incompressible by any recursive/computable procedure based on the chosen framework, if the observer only a priori knows the behavior of the nodes when isolated.
In other words, the emergent behavior is eventually\footnote{ As the network size increases.} non deducible---even in principle---from the parts for any above described external observer.
Thus, we say EEOE is an \emph{asymptotic observer-independent emergent phenomenon}.

If the observer is part of network that is displaying EEOE, such an unpredictability may be actually magnified, since the observer in this case would only know its own behavior (or maybe also its immediate neighbor's) when isolated.
More than new emergent irreducible information from other individuals in the network appearing to the node/observer, the networked behavior of the very node/observer would appear to itself as emergent with respect to the isolated case (or with respect to a previous initial stage where the respective network computing didn't start yet).
Within the abstract realm of algorithmic networks, future research on this \emph{reflexive emergence} of complexity (i.e., an emergence of complexity that arises from the comparison of the interacting behavior of a agent with the isolated behavior of the same agent) may be fruitful for investigating the presence of a process of algorithmic-informational autonomy \cite{Villalobos2018} as being emergent from the networking interaction with the environment (i.e., the rest of the algorithmic network in which the node/system is part of).

In both cases, i.e., either as a desirable or an undesirable emergent property, the investigation of network topological properties and local rules of interactions that are capable of triggering EEOE, such as in \cite{Abrahao2017publishednat,Abrahao2018publishednat,Abrahao2019nat}, seems to be a fruitful line of research in the intersection of complex systems science, theoretical computer science, complex networks theory, and information theory.

\footnotesize

\bibliographystyle{abntex2-num}
\bibliography{2.2.1-CompleteRefs-Felipe}

\end{document}